\documentclass[10pt, conference, compsocconf]{IEEEtran}

\usepackage[utf8]{inputenc} % allow utf-8 input
\usepackage[T1]{fontenc}    % use 8-bit T1 fonts
\usepackage{hyperref}       % hyperlinks
\usepackage{url}            % simple URL typesetting
\usepackage{booktabs}       % professional-quality tables
\usepackage{amsfonts}       % blackboard math symbols
\usepackage{nicefrac}       % compact symbols for 1/2, etc.
\usepackage{microtype}      % microtypography
\usepackage[pdftex]{graphicx}
\usepackage{pifont}
\newcommand{\ours}{\textsc{Mystiko}}

\usepackage{cite}

\usepackage[cmex10]{amsmath}

\usepackage{algorithmic}

\usepackage{array}

\usepackage{mdwmath}
\usepackage{mdwtab}

\usepackage{eqparbox}

\usepackage{subfigure}

\usepackage{stfloats}
\usepackage{comment}

\usepackage[disable]{todonotes}
\usepackage{multirow}

\begin{document}
%\listoftodos

\title{\ours : Cloud-Mediated, Private, Federated Gradient Descent}
%Maximizing Privacy Without Losing Accuracy}

% The \author macro works with any number of authors. There are two commands
% used to separate the names and addresses of multiple authors: \And and \AND.
%
% Using \And between authors leaves it to LaTeX to determine where to break the
% lines. Using \AND forces a line break at that point. So, if LaTeX puts 3 of 4
% authors names on the first line, and the last on the second line, try using
% \AND instead of \And before the third author name.

\author{\IEEEauthorblockN{K. R. Jayaram}
\IEEEauthorblockA{IBM Research AI}
\and
\IEEEauthorblockN{Archit Verma}
\IEEEauthorblockA{IBM Research AI}
\and
\IEEEauthorblockN{Ashish Verma}
\IEEEauthorblockA{IBM Research AI}
\and
\IEEEauthorblockN{Gegi Thomas}
\IEEEauthorblockA{IBM Research AI}
\and
\IEEEauthorblockN{Colin Sutcher-Shepard}
\IEEEauthorblockA{RPI \& IBM Research AI}
}

\maketitle

\begin{abstract}
%    Cloud computing has played a significant role in democratizing machine learning
%    by making large scale learning possible. 
    Federated learning enables multiple, distributed participants (potentially on
    different clouds) to collaborate
    and train machine/deep learning models by sharing parameters/gradients.
    However, sharing gradients, instead of centralizing data, may not be as private as one would expect.
    Reverse engineering attacks~\cite{frederickson1, frederickson2} on plaintext gradients 
    have been demonstrated to be practically feasible. Existing solutions for differentially 
    private federated learning, while promising, lead to less accurate models and require nontrivial
    hyperparameter tuning.
    
    In this paper, we examine the use of additive homomorphic encryption 
    (specifically the Paillier cipher)
    to design secure federated gradient descent techniques that (i) do not require addition
    of statistical noise or hyperparameter tuning, (ii) does not alter the final accuracy or
    utility of the final model, (iii) ensure that the plaintext model parameters/gradients
    of a participant are never revealed to any other participant or third party coordinator involved
    in the federated learning job, (iv) minimize the trust placed in any third party coordinator and 
    (v) are efficient, with minimal overhead, and cost effective.
    
\end{abstract}

\begin{IEEEkeywords}
federated learning, gradient descent, aggregator, additive homomorphic encryption
\end{IEEEkeywords}

\section{Introduction}~\label{sec:intro}

%\todo[inline]{todonotes package added. Use \ todo[inline] for comments}

%Although several geo-distributed
%learning algorithms and platforms have been proposed, and are probably used for
%extremely large datasets, they are designed to work under a single organizational 
%domain (e.g., a centralized service like Instagram, or across two data centers of company X). 

Some of the early success of distributed machine/deep learning (ML/DL) in several
application domains~\cite{ml-success-1,ml-success-2} has been in the context of 
massive centralized data collection, either at a single datacenter or a cloud service.
However, centralized data collection at a (third-party) cloud service 
can be incredibly privacy-invasive, and 
can expose organizations (customers of the cloud service) to large legal liability 
when there is a data breach.
This is especially true in the case of healthcare data, voice transcripts,
home cameras, financial transactions, etc.
Centralized data collection often results in ``loss of control''
over data once it is uploaded -- ``Is the cloud service using my data as promised? Is it actually
deleting my data when it claims to do so?''. Organizations that have not been convinced by
privacy violations and loss of control, have been forced by governmental regulations (like
HIPAA and GDPR~\cite{cloudcomputinglaw}) to restrict data sharing with third party services.

%Geo-distribution, by itself, 
%does not imply collaboration between entities and these techniques
%require data collection by a centralized ``service'' (e.g., a photo sharing 
%service like Instagram).
%For example, one of Facebook’s latest Detectron models~\cite{Detectron2018} for object detection 
%was trained on 3.5 billion images from Instagram.
%Examples from the health care domain include tumor
%identification from radiographs, analysis of heart rate and EKG data, etc. 
%Other examples include training a model to improve the results of speech recognition, 
%detecting fraud in credit card transactions and identifying faces in surveillance videos.

Federated learning (FL) aims to mitigate the three aforementioned issues,
while maintaining accuracy of ML/DL models. 
An entity in an FL job can be as
small as a smart phone/watch or as large as an organization with multiple 
data centers. An FL algorithm aims to train a ML/DL model,
e.g., a specific neural network model or an XGBoost model, on multiple entities, each 
with its own ``local'' dataset, without exchanging any data.
This results in multiple
``local models'', which are then combined (aggregated) by exchanging only parameters 
(e.g., the weights of the a neural network model). An FL algorithm
may use a central co-ordinator to collect parameters of all local models
for aggregation; or it may be a peer-to-peer algorithm (broadcast, overlay multicast, etc.)

%It involves collaborative learning across entities,
%where data never leaves said entity.  Typically, in such an
%algorithm, the bulk of training happens inside the entity, resulting  based on techniques like 
%random sampling, broadcast and multicast on a mesh or tree topology.

Exchanging parameters of the model was thought to provide privacy for free, and reduce
the amount of trust that a participant in a federated learning job has to place
in the co-ordinator or in other participants. But, researchers have 
demonstrated~\cite{frederickson1, frederickson2, shokri-membership-inference} through so-called \emph{model inversion}
and \emph{membership inference} attacks that parameters of a model can be used to reconstruct data from which they 
are derived. So the challenge in federated learning is to ensure the privacy of the 
model parameters and gradients. 

%One technique used for 
%privacy-preserving federated learning is addition of noise
%to parameters or their gradients by each participant, before they are sent out 
%for aggregation~\cite{diff-private-sgd}. 
It has been demonstrated~\cite{kamalika-diff-private-sgd, abadi-dl-dp}
that statistical noise, when added in specific ways to (clipped) gradients before being sent out for aggregation, can ensure corresponding levels differential privacy (recall that differential
privacy is a probabilistic guarantee). However, the same papers~\cite{kamalika-diff-private-sgd, abadi-dl-dp} also demonstrate
that achieving differential privacy 
results in non-trivial loss of model accuracy. Furthermore, increasing differential
privacy reduces accuracy. Also, for a specific level of differential privacy,
accuracy depends on careful hyperparameter selection (learning rate, batch size and gradient
clip values). This can be tricky because not all batch sizes which are generally "acceptable"
for training a specific neural net model in a non-private manner give the best accuracy in
a differentially private setting. Combined with learning rate and clipping thresholds, the state space
of hyperparameter exploration quickly becomes large.

%\todo[color=green,inline]{this is a bit vague.. can we bit more specific about what parameters and what is tricky? May be say that choice of privacy vs SGD hypermaraters?}

In this paper, we present \ours, a system for cloud-mediated, private federated
deep learning based on gradient descent. This paper makes the following technical contributions:

\begin{enumerate}

    \item A novel cryptographic key generation and sharing technique that uses additive homomorphic encryption
    (specifically the Paillier cipher~\cite{paillier-perf}) to \emph{maximize privacy}
    of federated gradient descent in training deep neural networks \emph{without any loss 
    of accuracy}.
    \item A novel protocol to arrange participants in a ring, and structure communication
    between them so that non-aggregated plaintext gradients are never revealed to any 
    participant or co-ordinator.
%    \todo[inline]{need to bring in the key-sharing novelty and no harm to accuracy}
    \item Extensions of the above technique to broadcast and All-Reduce protocols,
    for increased efficiency.
    
    \item A detailed empirical evaluation of \ours\ against differential privacy schemes (added statistical noise)
    and state-of-the-art secure multiparty computation protocols, using realistic datasets and 
    neural network models. We demonstrate performance improvements of up to 6.1$\times$ when compared to the 
    SPDZ protocol~\cite{spdz-csiro}.
    
\end{enumerate}

\section{Focus, Trust Model and Assumptions}

%\section{Overview of Our Approach}
 We focus on the scenario where each participant is convinced
of the benefits (improvements in accuracy, robustness, etc.) of federated learning.
%We note that convincing participants to collaborate, by projecting potential gains
%in accuracy due to federation by computing certain metrics on each participant's data
%is an open research problem.
%\todo[inline]{above lines may be redundant, we can skip focus part}
Each participant is convinced enough that it follows the steps of the federation protocol
and does not collude with the co-ordinator to break the protocol. 
But the participant may be curious about the data of others, and 
it may be in their interest to reverse-engineer the model parameters to try
and discover other participants' data. We also assume that the co-ordinator is honest, but 
curious with respect to individual participant's data.
The particpants
want to reduce the required amount of trust in the coordinator as much as possible.
Specifically, we assume that each participant does not attempt to poison or skew the 
global model by maliciously generating weights. 

%, and is chosen by consensus among
%the participants (e.g., with a leader election protocol~\cite{leaderelection}.\todo[inline]{can co-ordinator be one of the participants in this protocol, then we may run into issues}

%This trust model also includes the simpler
%case where participants are forbidden from sharing data or the model parameters derived
%from the data due to regulatory reasons (e.g., FedRAMP, EU data protection guidelines~\cite{cloudcomputinglaw}).
%\todo[inline]{the last sentence is incomplete. But do we need to make this assumption as we already said that participatns will follow the protocol?}

{\bf Examples/Use Cases:} This trust model is predominantly found in enterprise federated learning.
A multinational bank having branches in multiple countries and so regulated locally (BankA, BankA US, BankA UK, BankA India, etc.), where the bank
wants to learn a fraud detection model across data of all its subsidiaries, but 
the data can not be transfered to a central location due to governmental 
data sovereignty/jurisdiction laws~\cite{cloudcomputinglaw}. Each participant
here is a subsidiary with its national data center(s) and the coordinator 
might be located in a cloud platform or a global datacenter. Another example is a set of
hospitals who want to collaborate to train a tumor detection model; each hospital
is unable to trust the others and unwilling to trust and transfer data to a central service.
Another example is a cloud-hosted machine learning service (e.g., Azure ML) which has
multiple (competing) corporate clients which do not trust each other but have some level of
trust in the cloud service to facilitate and secure federated learning.

%\todo[color=green,inline]{Should this be part of this section as this is just background?}

\section{Background}

{\bf Gradient Descent~\cite{ruder2016overview}:} Throughout this paper, we focus on deep neural network training algorithms 
which rely on \emph{distributed} gradient descent (including optimized variants like Adam, Adagrad~\cite{ruder2016overview}, etc.).
We assume that there are $P$ parties, each learning on its own dataset.
It is well known that gradient descent is a popular optimization algorithm to minimize the loss for an objective function
$J(\theta)$ parameterized by a model's parameters $\theta \in \mathbf{R}^d$
by updating the parameters at every step ($\theta~=~\theta~-~\eta~\nabla_{\theta}~J(\theta)$)
The frequency of model
updates varies with the type of gradient descent: the gradient is computed on the whole dataset 
in the case of plain gradient descent, on each data item in the case of stochastic gradient descent (SGD)
and on a small batch of items in the case of mini-batch gradient descent. 

%Hence, the parameter update step, in distributed mini-batch gradient
%descent can be expressed formally as:

%\begin{equation}
%    \theta~=~\theta~-~\eta~\nabla_{\theta}~J(\theta;x^{(i:i+b)};y^{(i:i+b)})
%\end{equation}

%$\nabla_{\theta}~J(\theta)$ .\todo[inline]{give a reference of SGD/DL book and then this section can be compressed a lot} The learning rate $\eta$ 

%where $b$ is the batch size and $x$ and $y$ are attributes of the data items. For simplicity,
%$J(\theta;x^{(i:i+b)};y^{(i:i+b)})$ is generally written as $\nabla_{\theta}~J(\theta)$.

In a distributed setting, each participant computes $\nabla_{\theta}~J(\theta)$ on a batch 
from its dataset, and these gradient vectors are averaged before updating the model parameters $\theta$.
That is, either each participant broadcasts its gradients $\nabla_{\theta}~J(\theta)$ to all
other participants (using all-to-all broadcast or optimized all-reduce) or sends its gradients
to a central co-ordinator for averaging. The averaged gradients from iteration $t$ 
are then used to update
model parameters in iteration $t+1$ to compute the model parameters of iteration $t+1$.
%An alternative to broadcast or all-reduce is to use parameter servers or aggregator processes
%to do the averaging.

{\bf Additive Homomorphic Encryption:} Homomorphic encryption allows 
computation on ciphertexts, generating an encrypted result which, when decrypted, 
matches the result of the operations as if they had been performed on the plaintext~\cite{gentry-homomorphic}.  
Fully homomorphic encryption is expensive, both in terms
of encryption/decryption time as well as the size of the ciphertext. However, 
averaging gradient vectors requires only addition (division by the total number of participants
can be done before or after encrypted aggregation). Hence, we employ 
additive homomorphic encryption like the Paillier cryptosystem~\cite{paillier-perf}.
The Paillier cryptosystem  
is an asymmetric algorithm for public key cryptography. Given only the public key and the 
encryption of $m_1$ and $m_2$, one can compute the encryption of $m_1~+~m_2$.

%\todo[inline]{can we provide the algorithmic complexity of FHE and Paillier?}

\begin{comment}
and predicting the next word that is typed on a mobile phone to help people type more quickly. 
In both cases, it would be beneficial to directly train on data from a person 
instead of using text from Wikipedia, etc.. This would allow training a model on 
the same data distribution that is also used for making predictions. However, 
directly collecting this data is a terrible idea because it is extremely private. 
Users do not want to send everything they type or say to a cloud service.
\end{comment}

\section{\ours\ -- Overview}
\begin{comment}
\begin{figure}
    \centering
    \includegraphics[width=0.8\columnwidth]{figures/Mystiko-Architecture}
    \caption{\ours\ -- Architecture}
    \label{fig:arch}
\end{figure}
\end{comment}

\begin{figure}[htbp]
    \centering
    \includegraphics[width=\columnwidth]{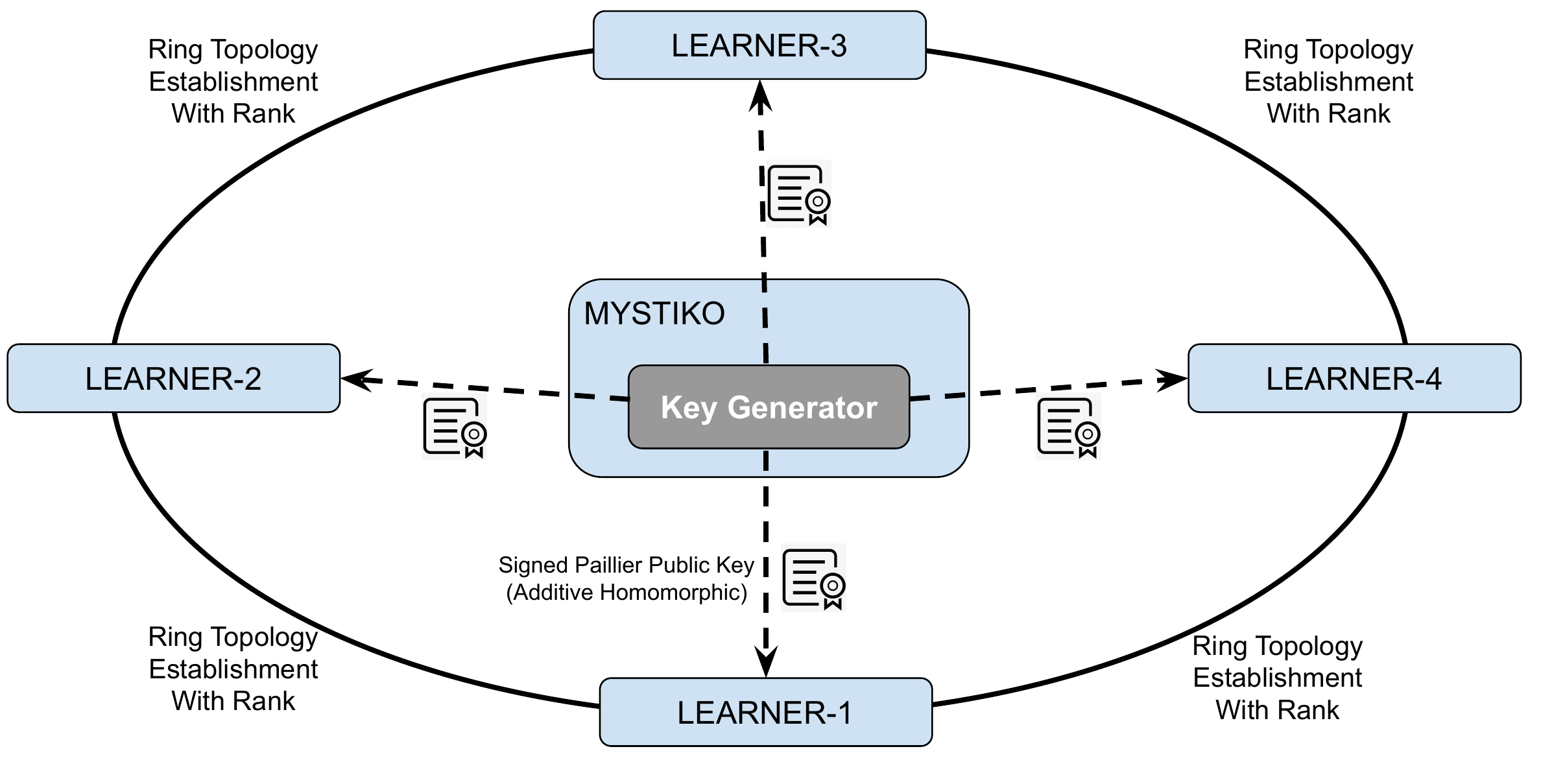}
    \caption{Topology establishment and Key Distribution}
    \label{fig:topology-key-distribution}
\end{figure}

\begin{figure}[htbp]
    \centering
    \includegraphics[width=\columnwidth]{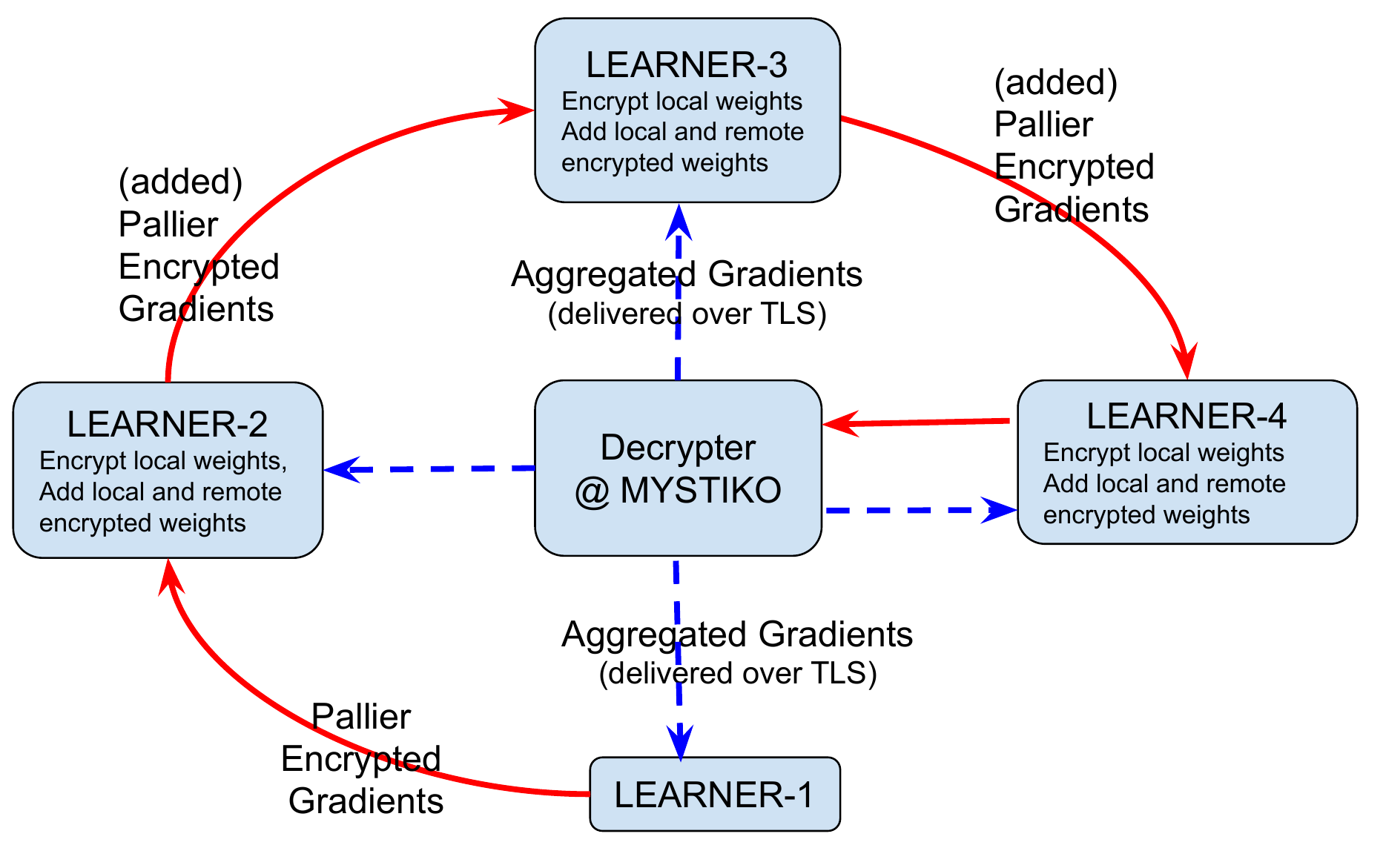}
    \caption{Basic Aggregation Protocol Over a Ring Topology}
    \label{fig:ring-aggregation}
\end{figure}

The main characteristics of any privacy preserving federated learning scheme revolves around a) what methodology is used to encrypt the data (or noise addition) of the participants and b) the communication protocol being used among the participants and the co-ordinator (if any) for aggregation of the model parameters or gradients. 
We present a novel approach, \ours\, in which all participants encrypt individual data using a single Paillier public encryption key, adding encrypted gradient vectors 
and decrypting only the sum. Thus, only the participants are able to view their
individual data, ensuring privacy. The question now is (1) how to distribute a 
common Paillier public key to all participants while keeping the corresponding private key
secret?, and  (2) how to prevent anyone from decrypting individual weights? These are explained in 
the following sections.

\subsection{Participants, Learners and Administrative Domains}

%In the enterprise case, federated learning typically crosses organizational 
%boundaries and often crosses geographical boundaries. 
Logically speaking,
it is helpful to define federated learning algorithms in terms of
\emph{administrative domains}. An administrative domain is a \emph{set of
computing entities} (servers, VMs, desktops, laptops, etc.). 
Each entity inside an administrative
domain trusts the other, malice is not a concern, and there are
no legal/regulatory hurdles to sharing data and information derived
from data. Note that an administrative domain does not necessarily mean
a company or a non-corporate organization. It may be a project within
a company handling confidential data; it may be not be located within
a corporate datacenter, and instead be associated with an account on the public cloud.
An organization can have multiple administrative domains. Each \emph{participant} in a
federated learning algorithm corresponds to an administrative domain. Computationally, the 
actual learning process (typically running on a GPU) performing the neural network
training is called a \emph{Learner}. Each Learner works on (a batch of) data within the participant
to compute the gradient vector.

\subsection{Architecture}

For simplicity, we assume that there is exactly one learner per participant. We will relax
this assumption in Section~\ref{sec:localagg}. \ours\ is typically deployed as a cloud
service that mediates several participants. It involves a Job Manager, Membership Manager, a 
Key Generator and a Decryptor. The Membership Manager is responsible for establishing 
the relationship between each participant and \ours, and also keeping track of
participants that belong to each federated learning job. The Job Manager manages an FL
job through its lifecycle -- it keeps track of participants, helps participants agree on
hyperparameters, detects failures and updates to memberships.

\subsection{Communication Security}
While the focus of this paper is attacks on the privacy of data from within
a federation, traditional communications security is nevertheless essential to 
prevent outside attacks on the federation.
For this, \ours\ and the participants (learners) 
agree to use a common public-key infrastructure (PKI)~\cite{nw-security-stallings}.
The PKI helps ensure confidentiality of communications between the \ours\ components and the 
learners, and helps bootstrap the Paillier infrastructure. 
The PKI provides certification authorities (CAs), along with corresponding intermediate and 
Root CAs, creating a web of trust between the learners and \ours. 
\ours\ creates a bidirectional TLS channel~\cite{tls-rfc-latest} using the PKI for 
the security of control messages. 
The TLS channel is created using strong, but ordinary (non-homomorphic) cryptographic algorithms
(e.g., RSA for key exchange/agreement and authentication, AES for message confidentiality and
SHA for message authentication~\cite{tls-rfc-latest, nw-security-stallings}). The TLS channel
is not used for gradient aggregation, but rather for all other communications, like registration
of learners with the \ours\, topology formation, rank assignment, transmitting the Paillier 
public key to each learner, and transmitting the decrypted aggregated gradient vector to each learner.

%\todo[color=green,inline]{shall we first keep it simple by having only learners in place of local aggregators before going to full local hierarchy?}

\section{Algorithms}~\label{sec:algos}

\subsection{Basic Ring-based Algorithm}~\label{sec:basicring}

The basic ring-based aggregation algorithm is illustrated in Figures~\ref{fig:topology-key-distribution} and \ref{fig:ring-aggregation}.
This algorithm operates across $P$ participants, each in its own administrative domain
and represented by a learner (L). The algorithm starts 
with each participant registering with \ours.
\ours\ acts as the coordinator.
The learners need not fully trust \ours; they only need 
to trust it to generate good encryption keypairs, keep the private key secret
and follow the protocol. 

\ours's Membership Manager starts the federated learning protocol once all expected learners
are registered. The first step is to arrange the learners along a ring topology (Figure~\ref{fig:topology-key-distribution}).
This can be done in several straightforward ways: (i) by location -- minimizing
geographic distance between participants, (ii) by following a hierarchy based on
the name of the participants (ascending or descending order), or (iii) by using
consistent hashing~\cite{consistenthashing} on the name/identity of the participants.
Once the learners have been arranged in a ring, each learner gets a rank 
(from 1 to $P$) based on its position in the ring. 

\ours's Paillier Key generator, which
generates a Paillier public- and private key pair for each federated learning job.
Typically, a unique Paillier key pair is generated for each federated learning job,
and the Paillier public key securely distributed (over TLS) to all the learners.
For long jobs, a separate keypair may be generated either once every epoch or once
every $h$ minutes (this is configurable). \ours's Decryptor is responsible for
decrypting the Paillier encrypted aggregated gradient vector for distribution to the learners.
 Each learner receives
a Paillier encrypted gradient vector from the previous learners on the ring, encrypts its
own gradient vector with the Paillier public key, and adds (aggregates) the two
Paillier encrypted vectors. This aggregated, Paillier encrypted gradient vector is then 
transmitted to the next learner on the ring. The last learner on the ring transmits the fully aggregated,
encrypted gradient vector to \ours\ for decryption. \ours's Decryptor decrypts the 
aggregated vector and transmits the same securely over TLS to each of the learners.

{\bf Security Analysis: } Data never leaves a learner, and by extension any administrative domain.
This ensures privacy of data, provided each server inside the administrative domain has
adequate defenses against intrusion. Unencrypted gradient vectors do not leave the learner.
If there are $P$ participants, in $P-1$ cases only aggregated Paillier encrypted gradient
vectors leave the learner. Only \ours\ has the private key to decrypt these. For the first 
learner in the ring, the non-aggregated gradient vector is transmitted to the second learner,
but it is Paillier encrypted, and cannot be decrypted by the same. In fact, none of the participants
are able to view even partially aggregated gradient vectors. After decryption, the aggregated gradient
vector is distributed securely to the participants over TLS. Reverse engineering attacks, like the 
ones in \cite{frederickson1} \& \cite{frederickson2} are intended to find the existence of a specific
data record in a participant, or to find data items that lead to specific changes in gradient vectors;
both of which are extremely difficult when several gradient vectors computed from large datasets are
averaged~\cite{bargav-dp}. Decryption after averaging ensures the privacy of gradients.
\ours\ only sees aggregated gradients and can't get access to individual learner's data or gradients.

%\todo[color=green,inline]{also mention that the TTP only sees aggregated encrypted gradients and so can't get access to individual LA's data}

{\bf Colluding Participants: } The basic ring-based algorithm is resistant to collusion among $P-2$ participants.
That is, assuming an honest uncorrupted \ours\ deployment, it can be broken only if $P-1$ learners collude.
For learner $L_i$'s gradients to be exposed, learners $L_1, L_2,\ldots, L_{i-1}$ and
$L_{i+1},\ldots,L_{P}$ should collude, i.e., all of them should simply pass on incoming 
encrypted gradient vector to the next learner, without adding any gradient of their own. 

%It can also be broken
%if a modified \mystiko\ colludes with the learner at the second position of the ring, and supplies
%$L_2$ with the private key for Paillier decryption. On the other hand, the ring based protocol
%is not broken if the TTP coordinator colludes with $LA_i$ for $i>2$, because the incoming gradient
%vector is already aggregated and individual gradient vectors can not be exposed.

{\bf Fault Tolerance:} The disadvantage of a ring-based aggregation algorithm is that rings
can break; for good performance, it is essential that the connectivity between each learner and 
\ours\ remains strong. Traditional failure detection techniques, based 
on heartbeats and estimation of typical round-trip times may be used. Distributed synchronous gradient 
descent consists of a number of iterations; with gradients being averaged at the end of
each iteration. If the failure of a learner is detected, the averaging of gradient vectors
is paused until the learner is eliminated from the ring by the \ours's Membership Manager, or connection to
the learner is established again. Pausing gradient averaging can also be done when connection to the
\ours\ is temporarily lost. 

\subsection{Broadcast Algorithm}

One of the main drawbacks of the ring based algorithm is the establishment and 
maintenance of the ring topology. To mitigate this, an alternative is to use 
group membership and broadcast. Except for the establishment of the topology,
the setting remains the same. Learners register with the \ours's Membership Manager; agree on 
a common PKI and know the identity and number of participants. \ours\ generates
a Paillier public-private key pair for each federated job and distributes the public key securely
to each learner.

\begin{figure*}[htbp]
\includegraphics[width=\textwidth,keepaspectratio=true]{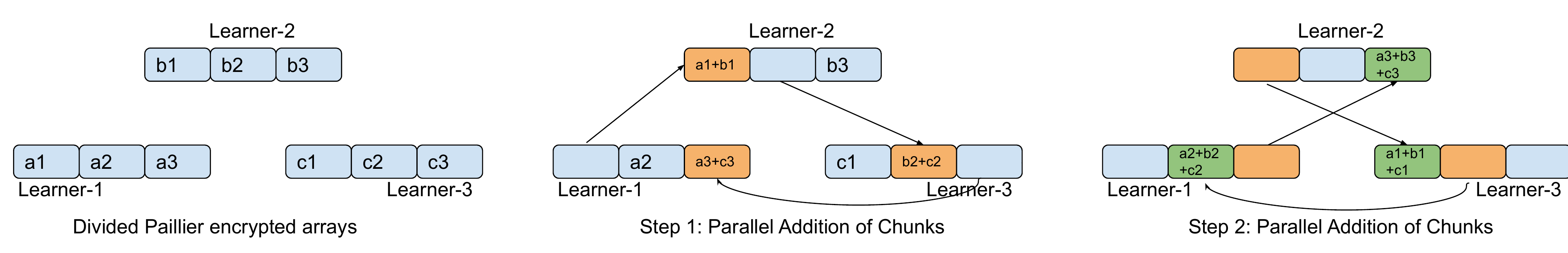}
\caption{\ours\ Ring All-Reduce over Paillier Encrypted Arrays}~\label{fig:all-red-illus}
\end{figure*}

\begin{figure}[htbp]
    \centering
    \includegraphics[width=\columnwidth]{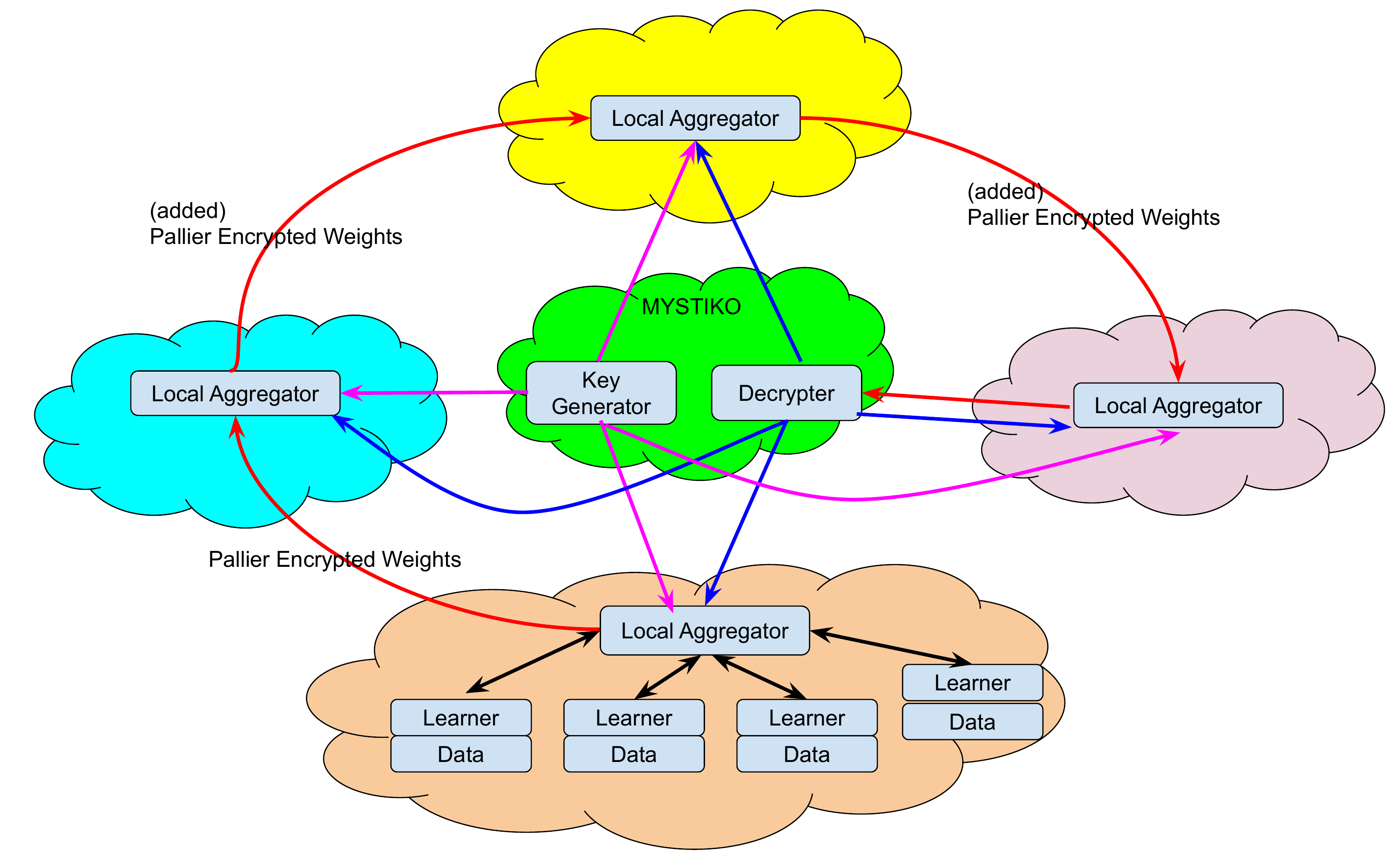}
    \caption{Federating Gradient Descent}
    \label{fig:mystiko-with-LA}
\end{figure}

Each learner Paillier-encrypts its gradient vector,
and broadcasts the encrypted vector to all other learners. Each learner, upon receipt of 
encrypted vectors from $P-1$ learnerss, adds them, and sends the Paillier-encrypted
sum to the \ours\ for decryption. After decryption, the aggregated gradient
vector is transmitted securely to all learners over TLS.
The broadcast algorithm is redundant and wasteful, as every
learner computes the aggregate. But, with redundancy comes increased failure resiliency.
With the ring, the failure of one participant can lead to partial loss of aggregated gradients,
which is not the case for broadcast. 
%\todo[color=green,inline]{isn't this quite wasteful as each LA is performing aggregation and sending to TTP while only one aggregation is sufficient? Can we think more like a tree, so that it only aggregates upto the root which sends to TTP?}

{\bf Colluding Participants: } The objective of breaking this algorithm is to 
determine the plaintext gradient vector of a specific LA. 
This algorithm is resistant to collusion, and can be
broken only if $P-1$ participants collude, which is highly unlikely. Also, in the event
that $P-1$ participants collude to Paillier encrypt zero-vectors instead of their actual
gradient vectors, the broadcasted Paillier ciphertexts from all the $P-1$ learners will be the same, which %\todo[inline]{didn't get this.. won't the aggregated vector from all will be same even if P-1 collude}
serves as a red flag enabling collusion detection. In fact, given that data is likely to be 
different at each participant, getting exactly the same Paillier encrypted gradient vector
from even two learners is red flag. 

%Consequently, to break this algorithm,
%each of the $P-1$ participants should agree, apriori, on the (fake) values of their gradient vectors,
%so that adding the $P-1$ vectors results in the zero vector. For example, with four participants, 
%three can agree to use the following (fake) vectors $[2,2,\ldots,2]$, $[5,5,\ldots,5]$ and 
%$[-7, -7, \ldots, -7]$. This agreement on the (fake) gradient vector values must happen for
%every iteration, over thousands of iterations, making attacks on this protocol
%increasingly difficult.

\subsection{All-Reduce}

Ring-based All-Reduce~\cite{ring-all-reduce} is essentially a parallel version of the
ring based aggregation protocol described in Section~\ref{sec:basicring}. It is illustrated
in Fig.~\ref{fig:all-red-illus}. The problem with the basic ring protocol in Section~\ref{sec:basicring}
is that each learner has to wait for its predecessor. However, in All-Reduce, the Paillier encrypted gradient 
vector is divided into $P$ chunks where $P$ is the number of participants. All learners
then aggregate Paillier encrypted chunks in parallel. For example, in Fig.~\ref{fig:all-red-illus}, there are
three learners, and the gradient vectors are divided into three chunks each. Learner-2 does not wait
for the entire vector of Learner-1 to be received. Instead, while it is receiving the first chunk of Learner-1,
it transmits its own second chunk to Learner-3, which in parallel, transmits its third chunk to Learner-3.
In Step 2, Learner-2 transmits the partially aggregated chunk-1 to Learner-3, which transmits partially
aggregated chunk-2 to Learner-1, which transmits the partially aggregated chunk-3 to Learner-2. At the end of 
Step-2, each learner has Paillier encrypted, aggregated chunks, which are transmitted to \ours's decryptor
for concatenation and decryption.

{\bf Security Analysis:} We note that All-Reduce is the most efficient \ours\ protocol.
With $P$ learners, All-Reduce is essentially an instantiation of $P-1$ rings (of Section~\ref{sec:basicring})
all operating in parallel. In Fig.~\ref{fig:all-red-illus}, the first ring starts at the 
first chunk of Learner-1, the second ring starts at the second chunk of Learner-2 and the
third starts at the third chunk of Learner-3. This implies that the security guarantees of 
All-Reduce are the same as that of the basic ring protocol.

\subsection{Multiple Learners Per Admin. Domain}~\label{sec:localagg}
For presentation simplicity, we have assumed that there is exactly
one learning process (learner) per participant. More realistically, 
within an administrative domain, data is partitioned among servers and 
multiple training processes (learners), which periodically synchronize their gradient vectors
using an aggregator process local to the administrative domain.
This is done for various reasons, including datasets being large, 
compute resources being cheap and available and the desire to reduce 
training time. \ours's protocols can be applied in a straightforward manner
to this case, with \ours's protocols running between Local Aggregators (LAs)
instead of between learners. Local aggregation is not Paillier encrypted,
and non-private because compute resources within an administrative domain are
trusted and can share even raw data. But aggregation between LAs follows \ours's protocols.
This is illustrated in Fig.~\ref{fig:mystiko-with-LA}.

%\begin{figure}[htbp]
%    \centering
%    \includegraphics[width=\columnwidth]{figures/AdministrativeDomain.pdf}
%    \caption{Administrative Domain}
%    \label{fig:admindomain}
%\end{figure}

\section{Empirical Evaluation}~\label{sec:eval}

\begin{figure*}[htbp]
\includegraphics[width=\textwidth,keepaspectratio=true]{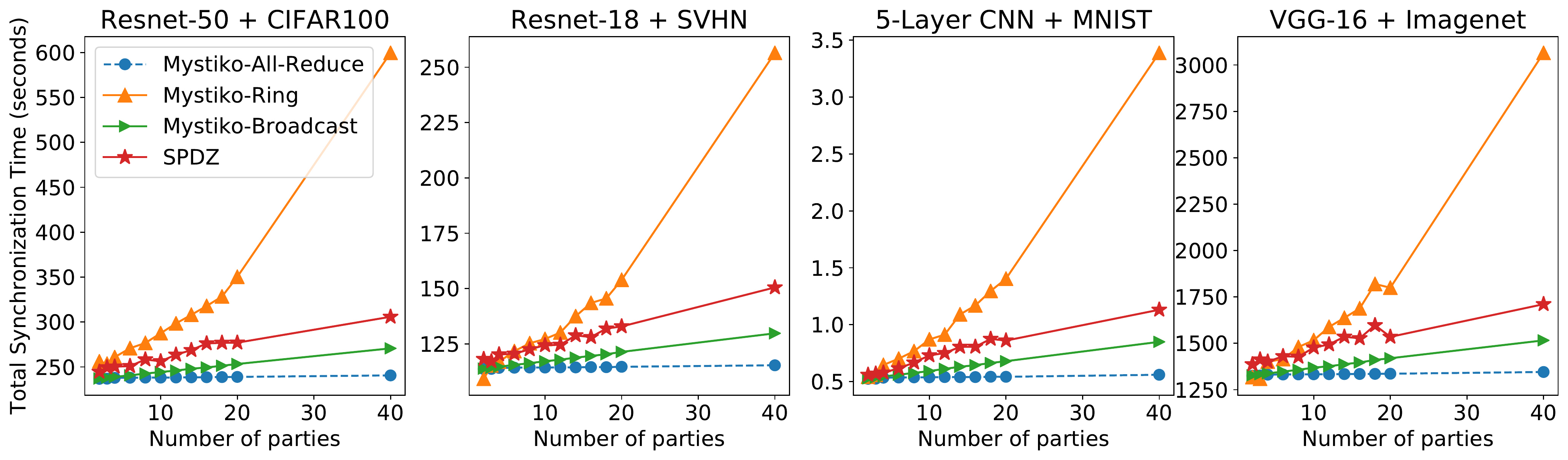}
\includegraphics[width=\textwidth,keepaspectratio=true]{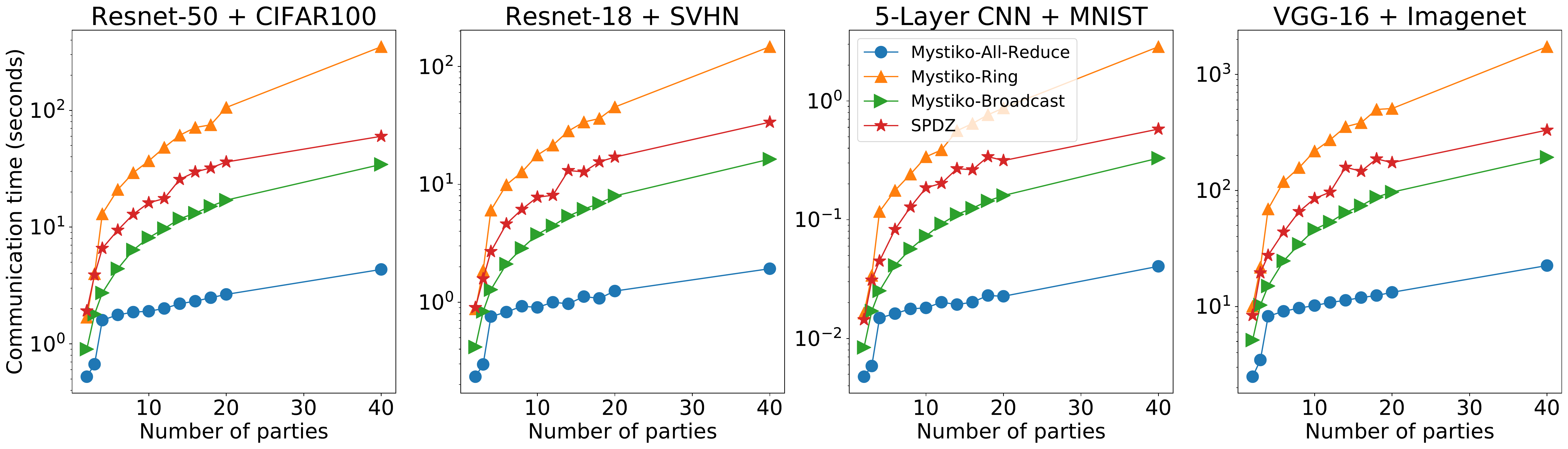}

\caption{Total Synchronization time (seconds) vs. Number of Parties (top plots) and Total communication time (seconds) vs. Number of Parties (bottom). Recall that Total Synchronization time is the sum of the communication time and the gradient transformation time.}~\label{fig:tot_synch_overhead}~\label{fig:comm_overhead}
\end{figure*}

\begin{table}[htbp]
    \centering
    \begin{tabular}{|c|c|c|c|c|}
    \hline
        \multicolumn{5}{|c|}{Communication Time}\\
        \hline
        & \multicolumn{3}{|c|}{\ours} & \\
        %\hline
       \# Parties & All Reduce  & Broadcast  & Ring & SPDZ  \\
       \hline
               20 & 1 & 6.4-7.2$\times$  & 38.2-39.9$\times$ & 13.1-14.2$\times$  \\

        40 & 1 & 7.9-8.6$\times$  & 70.5-80.8$\times$ & 13.8-14.7$\times$ \\
        \hline
        & & & & \\
        \multicolumn{5}{|c|}{Total synchronization time}\\
        \hline
        & \multicolumn{3}{|c|}{\ours} & \\
        %\hline
       \# Parties & All Reduce  & Broadcast  & Ring & SPDZ  \\
       \hline
               20 & 1 & 12-51\%  & 27-100\% & 2.2-6.1$\times$  \\

        40 & 1 & 6-25\%  & 15-59\% & 1.3-2.6$\times$ \\
        \hline
    \end{tabular}
    \caption{Performance slowdown of Broadcast, Ring and SPDZ relative to AllReduce. Full trend available at Fig.~\ref{fig:comm_overhead}}
    \label{tab:comm_sample}\label{tab:tot_sample}
\end{table}

\begin{table*}[htbp]
    \centering
    \begin{tabular}{|l|c|c|c|c|c|c|c|c|}
    \hline
           & \multicolumn{2}{|c|}{5-Layer CNN + MNIST} & \multicolumn{2}{|c|}{SVHN+Resnet-18} & \multicolumn{2}{|c|}{CIFAR100+Resnet-50} & \multicolumn{2}{|c|}{Imagenet+VGG-16} \\
           & & & & & & & & \\
           
    System & Synch. & Epoch & Synch. & Epoch & Synch. & Epoch & Synch. & Epoch \\
           & Time (s) & Time (s) & Time (s) & Time (s) & Time (s) & Time (s) & Time (s) & Time (s) \\
           \hline
           
           Non-private  & 0.03s & \multirow{5}{*}{0.71s} & 1.7s & \multirow{5}{*}{43.2s}& 3.7s & \multirow{5}{*}{184.6s} & 20.4s & \multirow{5}{*}{33490.5s}\\
           
           SPDZ  & 1.1s & & 150.5s & & 305.8s & & 1710.9s & \\ 

           \ours\ Ring  & 3.4s & & 256.5s & & 599.6s & & 3064.9 & \\

           \ours\ Broadcast & 0.84s  & & 129.8s & & 270.7ss & & 1515.7s & \\

           \ours\ All-Reduce  & 0.56s & & 115.3s  & & 240.6s & & 1345.5s & \\
           \hline
    \end{tabular}
    \caption{Synchronization time vs. Training time (for 1 epoch on 1 V100 GPU). 40 participant case.}~\label{tab:epoch-vs-synch}
\end{table*}

\begin{figure}[htbp]
    \centering
    \includegraphics[width=\columnwidth, keepaspectratio=true]{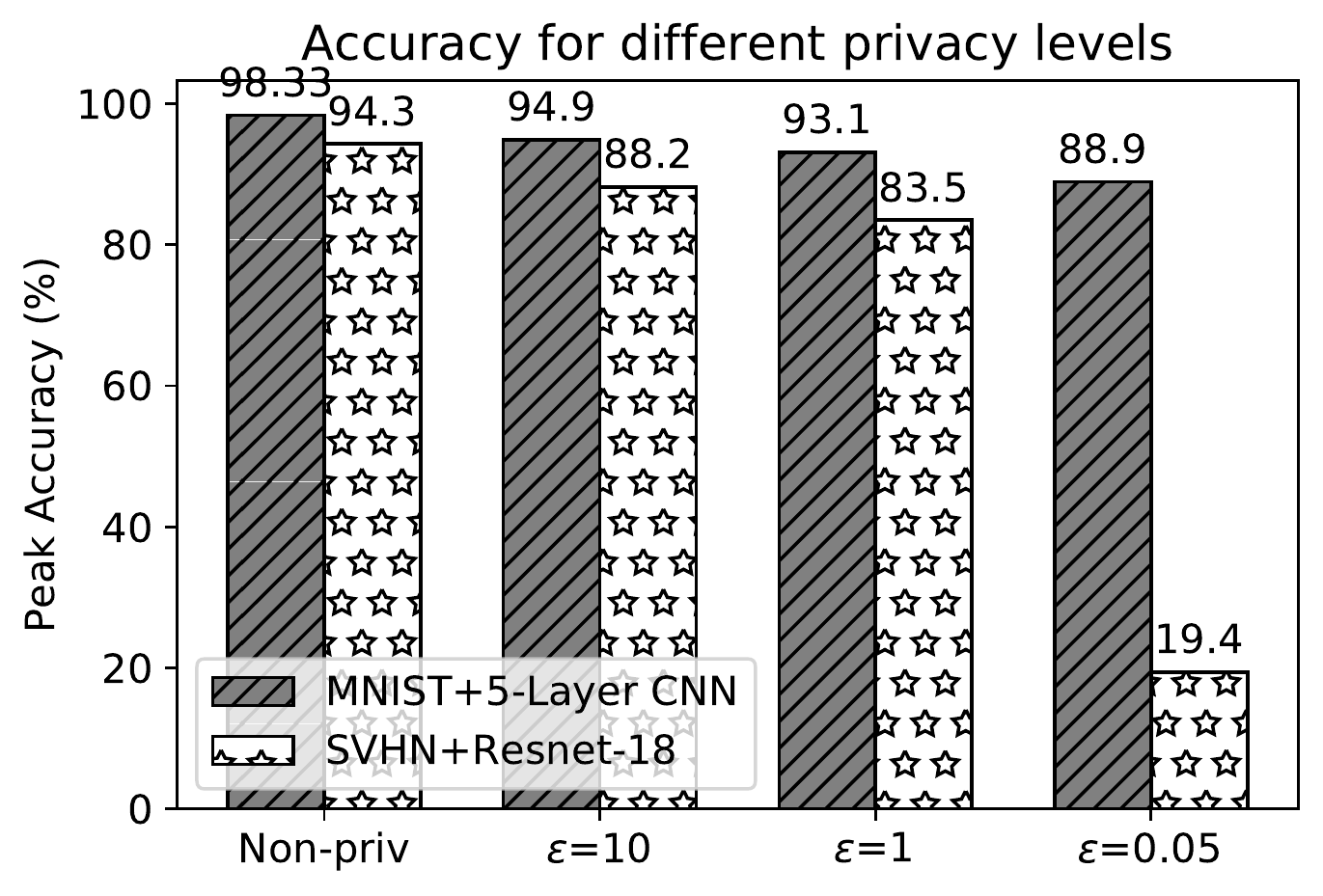}
    \caption{Reduction in validation accuracy when differentially private noise~\cite{abadi-dl-dp,kamalika-diff-private-sgd} is added.
    \ours\ and SPDZ do not add noise, and hence always achieves peak accuracy.}
    \label{fig:dp-accuracy-loss}
\end{figure}

We compare all the three Mystiko algorithms with a state of the art protocol for
secure multi-party computation -- SPDZ~\cite{spdz-bmr-eurocrypt, damgard-spdz, spdz-csiro}, 
and schemes for differential privacy (DP) through the addition of statistical noise.

\subsection{Implementation}
We have implemented \ours\ and all its protocols in Python and C++, using appropriate libraries
for PKI and communications security. We use the ETCD~\cite{etcd}
key value store to store metadata about federated jobs, and ETCD
leases for failure detection of learners.
We have implemented an optimized version of the Paillier cipher
following the improvements outlined in \cite{paillier-perf}.
In all our experiments, we use PyTorch 
for training on the learner side.
We have also integrated PyTorch on the learner side to work with the implementation of SPDZ2k from MP-SPDZ~\cite{spdz-csiro}
for gradient synchronization..

\subsection{Models and Datasets}
We employ a variety of image processing neural network models and datasets of
various sizes: (i) 5-Layer CNN (small, 1MB) trained on MNIST dataset (60K handwritten digit images)
(ii) Resnet-18 (small-medium, 50MB) trained on the SVHN dataset (600K street digit images),
(iii) Resnet-50 (medium-sized model, 110 MB) trained on CIFAR-100 dataset (60K color images of 100 classes) and
(iv) VGG-16 (large model, 600MB) trained on Imagenet-1K dataset (14.2 million images of 1000 classes).

\subsection{Experimental Setup}
Experiments were conducted on a 40 machine cluster to evaluate all the algorithms
on a varying number of participants from 2 to 40. No more than one participant
was ever run on any machine, each of which was equipped with 8 Intel Xeon E5-4110 (2.10 GHz)
cores, 64GB RAM, 1 NVIDIA V100 GPU and a 10GbE network link. The machines were spread over four datacenters, and in every experiment,
participants were uniformly distributed across datacenters. In every experiment, the dataset 
was \emph{uniformly and randomly} partitioned across participants. Mystiko was executed on a dedicated machine in one datacenter.
All data points henceforth are computed by averaging 10 experiment runs.

\subsection{\ours\ vs. DP Noise Addition~\cite{abadi-dl-dp, kamalika-diff-private-sgd, bargav-dp}}

It has been already demonstrated~\cite{bargav-dp} that while addition of differentially private noise
mitigates membership inference~\cite{shokri-membership-inference} and reverse engineering~\cite{frederickson1,frederickson2}
attacks, the resulting model has significantly lower accuracy as illustrated in Fig.~\ref{fig:dp-accuracy-loss}.
In Fig.~\ref{fig:dp-accuracy-loss}, the differential privacy parameter is $\epsilon$, and the
peak accuracy for a given $\epsilon$ value corresponds to one achieved by trying different 
combinations of learning rate and batch size. Acceptable levels of differential privacy
are achieved for $\epsilon~\leq~1$, and values less than 0.1 are preferred for neural network
training because gradients are exposed over thousands of synchronization steps. We observe from Fig.~\ref{fig:dp-accuracy-loss}
that even for simple datasets and neural network models, loss of accuracy is significant.
Noise addition is also not computationally cheap (298.5s vs. 43.6s per epoch in the case of SVHN+Resnet-18).

%\begin{figure*}[htbp]
%    \includegraphics[width=\textwidth,keepaspectratio=true]{graphs/comm_all.pdf}
%    \caption{Communications time (seconds) vs. Number of Parties. Y-axis is LOG-SCALE.}
%    \label{fig:comm_overhead}
%\end{figure*}

\subsection{Comparing \ours\ and SPDZ}
In federated learning, learners (or local aggregators) learn on local data for a specific
number of iterations before federated gradient aggregation and model update. 
Privacy loss happens during gradient aggregation, which is where \ours\ and other 
systems like SPDZ intervene. Hence, we use the following two metrics to evaluate 
\ours\ and SPDZ: (i) Total synchronization time, which measures the total time needed
for privacy preserving gradient transformations (Paillier encryption in \ours, share generation in SPDZ, etc.)
and the time required to communicate the transformed gradients to participants for federation,
and (ii) communication time, which only measures communication time. 

Figures~\ref{fig:tot_synch_overhead} plots total synchronization time 
and communication time against the number for parties involved in federation, for all of our model/dataset
combinations. From Fig.~\ref{fig:comm_overhead}, we observe that All-Reduce is the most scalable of 
all the protocols, as the number of participants increases. This is mainly because it is a parallel protocol,
where each learner/LA is constantly transmitting a small portion of the gradient array. 
The basic ring protocol is the least scalable, because it is sequential. Broadcast performs and scales
better than the basic ring protocol because each participant is broadcasting without waiting for the others. 
SPDZ performs and scales worse than broadcast because its communication pattern is close to (but not exactly)
a dual broadcast -- each item of the gradient vector at each participant is split into secret shares and broadcast
to the other participants; after secure aggregation, the results are broadcast back. \ours\ obviates the
need for dual broadcast through the use of Paillier encryption and centralized decryption of aggregated
gradients. Table~\ref{tab:comm_sample} illustrates the performance impact of using other protocols for 
two cases (20 and 40 parties from Fig.~\ref{fig:comm_overhead}).

However, the ``enormous'' speedups of using All-Reduce do not materialize in the total synchronization
reduce when total synchronization time is considered. The scalability trends among the four protocols
remain the same; the speedups in total synchronization time \emph{remain significant} (as elucidated for two
cases in Table~\ref{tab:tot_sample}). But the speedups are lower than the speedups due to communication.
This demonstrates that the predominant overhead of private gradient descent in \ours\ and SPDZ vs.
non-private gradient descent is gradient transformation prior to communication.
From Fig.~\ref{fig:tot_synch_overhead} and Table~\ref{tab:comm_sample}, we also observe that for small models
(5-Layer CNN \& Resnet-18), communication time plays a larger role. But for large models (Resnet-50 and VGG-16),
gradient transformation plays a larger role.

Lastly, we observe from Table~\ref{tab:epoch-vs-synch} that when compared to training time (illustrated
using epoch time), synchronization time for private gradient descent is significantly larger than non-private
gradient descent. This is primarily because training happens on V100 GPUs (with thousands of cores) while
gradient transformation happens on CPUs. While there is a GPU accelerated version of fully homomorphic encryption
(\cite{cuhe} which has worse performance than Paillier on CPUs), we are not aware of any GPU accelerated 
version of the Paillier algorithm.

\section{Related Work}~\label{sec:related}

\begin{table*}[htbp]
\centering
\begin{tabular}{|c|c|c|c|c|}

\hline
   Work   & Accuracy & Synch. & Min \# of & Peers  assumed to be \\
                &          & Overhead    & Colluding parties  &  \\
                &          &             & (to violate privacy) & \\ 
\hline
Aono et.al.~\cite{aono-dl-homomorphic} & Unch. & Moderate & 1 & Trusted \\
 & & (ENC Only) & & \\
\hline
Diff. Priv.                       &           &          &     & Honest  \\
Abadi et.al.~\cite{abadi-dl-dp}         & Lower     & Moderate      & $P-1$ & curious \\
Song et.al.~\cite{kamalika-diff-private-sgd}  & & (NA only) &  & \\
\hline
%Hom. enc. + Diff. Priv.  & & &  &\\
HybridAlpha~\cite{nathalie-hybridalpha}  & Lower & High & $P-1$ & Honest \\

Turex et.al.~\cite{nathalie-hybrid-eff}  & Lower & High & $P-t-1$ & Curious \\ 

\hline
%Classic SMC & & & & & & \\
%\cite{smc-book, evans-smc-pragmatic} & \cmark & \cmark & Unchanged & High? & ?  \\
%\hline
SPDZ variants  & & & &\\
\cite{damgard-spdz,spdz-bmr-eurocrypt}   & Unch. & High & P-1  & Honest \\
\cite{overdrive, sahai-smc} & & (Sec.~\ref{sec:eval}) & & Curious \\
\hline
\ours\  & Unch. & High & P-1  & Honest\\
Ring, Broadcast & & (Sec.~\ref{sec:eval}) & & Curious\\
\hline
\ours\  & Unch. & Moderate & P-1 & Honest \\ 
All-Reduce & & (Sec.~\ref{sec:eval}) & & Curious\\
\hline

\end{tabular}

\caption{Comparing \ours\ with the state of the art. ENC means Paillier Encryption, NA means Differential privacy through noise addition. Based on the experiments in Section~\ref{sec:eval}, we relatively characterize the overhead of 
Paillier encryption with All-Reduce as ``Moderate'', Noise Addition with All-Reduce as ``Moderate'', and combinations of both encryption and noise addition as ``High''}~\label{tab:relatedwork}
\end{table*}

%\todo[inline]{TableIV:what's the crieterion used to say low/moderate/high for computation overhead?  can we expand the caption to make it as self-contained as possible?}

Research on secure and private federated learning and gradient descent is predominantly
based either on (1) clever use of cryptography -- homomorphic encryption and 
secure multi-party computation~\cite{aono-logistic-homomorphic, aono-dl-homomorphic, smc-book, damgard-spdz, sahai-smc, overdrive, evans-smc-pragmatic, spdz-bmr-eurocrypt}, or on (2) modifying model parameters or gradients
through the addition of statistical noise to get differential privacy~\cite{abadi-dl-dp, kamalika-diff-private-sgd, cpsgd}. Some techniques~\cite{nathalie-hybridalpha, nathalie-hybrid-eff}
combine both.

We compare our proposed algorithms with other techniques in Table~\ref{tab:relatedwork}.
We observe that no technique is perfect -- each has pros and cons, which must be 
carefully considered based on the deployment environment.
We examine whether each algorithm (i) requires fully-trusted peers or whether it can
work with honest+curious peers, (ii) impact on final model
validation accuracy, (iii) synchronization overheads (due to encryption and other aspects of the 
algorithms), and (iv) number of participants that must collude to violate the privacy guarantees 
of the protocol (assuming $P$ total participants).

The closest related work to our algorithms is from Aono et. al.~\cite{aono-logistic-homomorphic, aono-dl-homomorphic}, where the participants jointly generate a Paillier-keypair;
send the encrypted gradient vectors to the coordinator who is completely untrusted,
except to add Paillier encrypted weights. The participants can then decrypt the aggregated
gradient vectors. This, however, requires each participant to collaborate with the 
others to generate the Paillier keys and a high level of trust that participants do
not collude with the untrusted coordinator to decrypt individual gradient vectors.
One untrusted participant can leak the Paillier keys and potentially lead to privacy loss.

Adding statistical noise to gradients has been proposed in ~\cite{kamalika-diff-private-sgd, abadi-dl-dp, cpsgd} among other literature~\cite{nathalie-hybrid-eff,nathalie-hybridalpha}. If noise is added in a manner that guarantees 
differential privacy of the gradient vectors, the gradients can be released 
publicly to the other participants and the public. In this sense, these techniques
work with any type of participant or coordinator. Although computation overhead is moderate,
accuracy of the final model takes a hit as demonstrated in Section~\ref{sec:eval}, and these techniques require careful
hyperparameter selection to keep the loss of accuracy minimal.

Secure multi-party computation (SMC) is a subfield of cryptography with the goal of creating 
methods for parties to jointly compute a function over their inputs while keeping those 
inputs private. Unlike traditional cryptographic tasks, where cryptography assures security 
and integrity of communication or storage and the adversary is outside the system of 
participants (an eavesdropper on the sender and receiver), the cryptography in this model protects participants' privacy from each other. SPDZ~\cite{damgard-spdz}, and its
variants (Overdrive~\cite{overdrive},\cite{spdz-bmr-eurocrypt}, \cite{sahai-smc})
optimizes classic SMC protocols. The advantage of such protocols is 
that they work with any number of honest+curious peers, do not
change final accuracy of the trained model, and require a large number of colluding 
peers to break. The drawback, however, is efficiency -- SMC protocols
are computationally expensive (Section~\ref{sec:eval}).

%\todo[inline]{what do check-marks and cross mean in the first two columns in table II? Are they honest+curious or not? WHat is the difference between Classic SMC and SPDZ rows? Do we need two separate rows? Need to complete this for third rows, hybridalpha etc.}

\section{Conclusions and Future Work}~\label{sec:conclusions}
Through \ours, we have demonstrated that private, federated gradient descent without
loss of accuracy can be practical with reasonable synchronization overhead. 
\ours\ is the first federated learning platform to provide the same 
security and privacy guarantees of a state-of-the-art secure multiparty
computation protocol (MP-SPDZ~\cite{spdz-csiro}) with significantly
lower synchronization overhead and by using only Paillier encryption.
Another important characteristic of \ours\ is that it is simple
to understand, implement and explain to users (vis-a-vis protocols like SPDZ).
However, we do observe that synchronization time remains large
with respect to epoch time for many models. We believe that this can
be mitigated by effectively accelerating Paillier encryption using GPUs, and this is 
a topic of active research in our organization.

\bibliographystyle{IEEETran}
\bibliography{privfedsgd}

\end{document}